\documentclass[12pt]{iopart}

\expandafter\let\csname equation*\endcsname\relax
\expandafter\let\csname endequation*\endcsname\relax

\usepackage{amsmath}
\usepackage{mathtools}
\usepackage{cite}
\usepackage[page]{appendix}
 \usepackage{iopams}
\usepackage{graphicx}
\usepackage{float}
\usepackage{xcolor}
\usepackage{epstopdf}
 \usepackage{appendix}
 \usepackage{geometry}

\newcommand{\ket}[1]{\bigl| #1 \bigr\rangle}
\begin{document}
\title[]{Entangling power of  Anti-Jaynes-Cummings model and its efficiency to encode information in atomic system }
\author{M. Kh. Ismail$^{1}$,T. M. El-Shahat$^{1}$, N. Metwally$^{2,3}$ and A.-S. F. Obada$^{4}$}
\address{$^{1}$ Math. Dept.Faculty of Science, Al-Azhar University, Assiut 71524, Egypt.
\\
$^{2}$ Math. Dept, Faculty of Science, Aswan University, Aswan, Egypt.\\
$^{3}$ Department of mathematics, college of science, university of Bahrain, Bahrain.\\
         $^{4}$ Math. Dept. Faculty of Science, Al-Azhar University, Nasr City 11884, Cairo, Egypt.
}

\begin{abstract}

Entangling power is crucial for quantum information processing. This study examines the Anti-Jaynes-Cummings Model (AJCM) in generating quantum correlations between two atoms interacting via the Ising model and its effect on the entangled system. The AJCM is shown to create entanglement suitable as a quantum channel for information encoding. Interaction parameters act as controls to enhance quantum correlations, increase the capacity of the final atomic state, and improve system efficiency. When the atomic system starts in a maximally entangled state, increasing interaction strength and mean photon number further boosts concurrence and channel capacity.

\end{abstract}

\section{Introduction}

Quantum computation and quantum information \cite{RR1,RR2,RR3} have become highly active research fields due to their potential to surpass classical systems. As a fundamental resource, quantum entanglement plays a crucial role in quantum information processing tasks such as teleportation \cite{Metwally}, key distribution, and dense coding \cite{RR4,RR5,RR6,Metwally1}. Significant efforts have focused on generating, quantifying, and enhancing entanglement. Regarding entanglement generation, a key question is the capability of a given operation to produce entanglement. Recently, studies have explored the entanglement capabilities of quantum evolution and Hamiltonians \cite{RR7,RR8,RR9,RR10}..

In general, we can apply unitary operators to pure product states and examine the amount of entanglement generated, which always depends on the input states. Two methods have been proposed to obtain an input-state-independent measure of a unitary operator's entanglement creation capacity. One approach is to take the maximum entanglement produced over all product input states \cite{RR11,RR12}, while the other averages over the product input states \cite{RR13,RR14,RR15}. The entangling power falls into the second category \cite{RR16,RR17}. Additionally, several studies have shown that two isolated qubits can become entangled when interacting with a common heat bath, indicating that even noise-induced evolution may possess entangling power. Based on completely positive (CP) maps, Zanardi et al  \cite{RR15} defined the partial entangling power of a unitary transformation, and the partial entangling power of the Jaynes-Cummings model has also been explored \cite{RR18}. For unitary quantum gates, entangling power has been extended to many-body systems  \cite{RR19,RR20}.

On the other hand, quantum dense coding (QDC) has become an interesting topic in quantum computation and quantum information \cite{RR21,RR22}.
As one of the essential applications of quantum entanglement.
QDC can transmit more classical information by passing on less quantum resources. The key feature of dense coding protocol is that one qubit of entanglement allows Alice to send two classical bits of information to Bob. This is because the entangled pair initially has more information (due to their entanglement) than each qubit individually. From then on, piles of works on dense coding have been introduced theoretically \cite{RR23,RR24}, and experimentally \cite{RR25}.
\\
Various systems have been explored in quantum processes, with spin chains studied as promising candidates for quantum correlations and entanglement \cite{RR26,RR27}. Additionally, the cavity quantum electrodynamics system (CQES) has gained significant attention due to its optimal coupling between atoms and photons \cite{RR28}. CQES has been widely used to engineer quantum entanglement and quantum channels \cite{RR29,RR30,RR31,RR32,RR33}, and is not only effective for entanglement generation but also for applications like quantum state transfer. These systems are also well-suited for dense coding.

The development of various quantum technology applications depends on understanding and analyzing interactions between quantum fields and atomic systems. A system of two-level atoms interacting with a quantized electromagnetic field and with each other through an Ising-like interaction provides a fundamental example. Traditionally, the Jaynes-Cummings Model (JCM) has been used to theoretically describe such systems \cite{RR034,RR035,RR036,RR037,RR038}. However, growing interest has recently been directed toward studying field-atom interactions through the Anti-Jaynes-Cummings Model (AJCM) \cite{RR039,RR040,RR041}. The influence of atom-cavity interactions and artificial magnetic fields on quantum phase transitions in the AJCM triangle model, particularly in the infinite frequency limit, has been explored \cite{RR042}. Additionally, the supersymmetric connection between the JCM and AJCM in quantum optics has been studied \cite{RR043}. The ability of both models to generate quantum correlations between finite- and infinite-dimensional subsystems has also been investigated \cite{RR18,RR044}. Notably, it has been demonstrated that the AJCM generally produces entanglement more rapidly and achieves greater entangling power within a shorter time frame \cite{Ivan}

Therefore, we are motivated  to examine the ability of the AJCM to generate entanglement  between two atoms initially prepared in separable states. 
Moreover, we aim to investigate whether this amount of entanglement is sufficient to be used as a quantum channel to encode information.
This paper is organized as follows. In Sec.$ 2$, the model  and its solution are introduced. The impact of the interaction parameter on the entangling power is discussed in Sec.$3$. In Sec.$4$, the concurrence is used as a quantifier of the amount of entanglement contained in the final atomic state. The behavior of the channel capacity is investigated in Sec.$5$. Finally our results are summarized in Sec.$6$.

\section{ The Hamiltonian system and quantum dynamics}
Assume that we have two-level identical atoms interacting locally  with a cavity mode through the AJCM described by the Hamiltonian.
\begin{eqnarray}
  H_{I}&=&\lambda\sum_{l=1}^{2}\left(a^{\dagger}\sigma_{+}^{l}+a\sigma_{-}^{l}\right),
\end{eqnarray}
where $\lambda_{1}=\lambda_{2}=\lambda$ are the atoms-field coupling strength and $\sigma_{+}^{l}=(\sigma_{-}^{l})^{\dag}=|e\rangle^{l}\langle g|$ the Pauli operators for the atom $l$ while $a$ and $a^{\dagger}$ denote the annihilation and creation operators. Moreover, there is an atom-atom interaction described by
\begin{eqnarray}
  H_{a-a}&=\Omega_{s}\sigma_{z}^{1}\sigma_{z}^{2},
\end{eqnarray}
where $\Omega_{s}$ is the interaction strength, $\sigma_{z}^{l}=|e\rangle^{l}\langle e|-|g\rangle^{l}\langle g|$ for the atom $l$. Then, the total interaction Hamiltonian which describes the whole system is given by,
\begin{eqnarray}
  H&=H_{I}+H_{a-a}.
\end{eqnarray}
Now, let us assume that the atomic system is initially prepared in the state, $\rho_{aa}=\cos^{2}(\theta)|ee\rangle\langle ee|+\sin^{2}(\theta)|gg\rangle\langle gg|+\sin(\theta)\cos(\theta)|ee\rangle\langle gg|+\sin(\theta)\cos(\theta)|gg\rangle\langle ee|$, $\theta$ $\in$ $[0,\pi]$ and the cavity mode is represented in its coherent state, $\rho^{in}=\sum_{m,n}C_{n,m}|m\rangle\langle n|$, where $\bar{n}$  is the mean photon, and $C_{m,n}=\frac{\bar{n}^{(n+m)/2}}{\sqrt{n!m!}}e^{-\bar{n}}$.
Therefore,  the initial state that describes the system is $\rho_s(0)=\rho^{in}(0)\otimes\rho_{aa}(0)$. At $t>0$,  the density operator of the system  can be written as
\begin{eqnarray}
\rho_s(t)=\mathcal{U}_{I}(t)(\rho^{in}(0)\otimes\rho_{aa}(0))\mathcal{U}_{I}^{\dagger}(t),
\end{eqnarray}
 where $\mathcal{U}_I(t)=exp\left(-iHt\right)$ is the time evolution operator. In the atomic basis set, $\{|ee\rangle,|eg\rangle,|ge\rangle,\textcolor{blue}{|gg\rangle}\}$  the unitary operator $\mathcal{U}_I(t)$  could be described by $4\times 4$ elements. The explicit mathematical forms of these elements are given by,
\begin{eqnarray}
 \nonumber
  \hat{U}_{11}&=&C_{\lambda_{s}}+2 a^{\dag}\frac{C_{M\lambda_s}}{f_{1}^2}a-i S_{00},
  \quad \hat{U}_{12}=-i a^{\dag}\frac{S_{M}}{M},\quad \hat{U}_{13}=-i a^{\dag}\frac{S_{M}}{M},
  \nonumber\\
   \hat{U}_{14}&=&
  2 a^{\dag}\frac{C_{M\lambda_s}}{f_{1}^2}a-iS_{03},~
  \hat{U}_{21}=-i\frac{S_{M}}{M}a,
  \quad \hat{U}_{22}=C_{\lambda_{s}}+\frac{C_{M\lambda_s}}{2}-i S_{11},\quad
  \nonumber\\
  \hat{U}_{23}&=&\frac{C_{M\lambda_s}}{2}-i S_{12},
  \hat{U}_{24}=-i\frac{S_{M}}{M}a^{\dag},~
  \hat{U}_{31}=-i\frac{S_{M}}{M}a,
  \nonumber\\
  \quad \hat{U}_{32}&=&\frac{C_{M\lambda_s}}{2}-i S_{21},\quad \hat{U}_{33}=C_{\lambda_{s}}+\frac{C_{M\lambda_s}}{2}-i S_{22},\quad \hat{U}_{34}=-i\frac{S_{M}}{M}a^{\dag},
  \nonumber\\
  \hat{U}_{41}&=&2a\frac{C_{M\lambda_s}}{f_{1}^2}a-iS_{30},
  \quad \hat{U}_{42}=-i a \frac{S_{M}}{M} ,\quad \hat{U}_{43}=-i a \frac{S_{M}}{M},\quad
  \nonumber\\
\hat{U}_{44}&=&C_{\lambda_{s}}+2 a\frac{C_{M\lambda_s}}{f_{1}^2}a^{\dag}-i S_{33},
\end{eqnarray}
where
\begin{eqnarray}
  S_{00} &=&S_{\lambda_{s}}-2a^{\dag}\left(\frac{S_{\lambda_{s}}}{f_{1}^2}-\lambda_{s}\frac{S_{M}}{Mf_{1}^2}\right)a,\qquad S_{03}=2a^{\dag}\left(\frac{S_{\lambda_{s}}}{f_{1}^2}-\lambda_{s}\frac{S_{M}}{Mf_{1}^2}\right)a^{\dag}
  \\\nonumber
  S_{11} &=&S_{22}=\frac{-1}{2}\left(S_{\lambda_{s}}+\lambda_{s}\frac{S_{M}}{M}\right),\qquad S_{12}=S_{21}=\frac{1}{2}\left(S_{\lambda_{s}}-\lambda_{s}\frac{S_{M}}{M}\right),
  \\\nonumber
  S_{30} &=&-2a\left(\frac{S_{\lambda_{s}}}{f_{1}^2}-\lambda_{s}\frac{S_{M}}{Mf_{1}^2}\right)a,\qquad S_{33}=S_{\lambda_{s}}-2a\left(\frac{S_{\lambda_{s}}}{f_{1}^2}-\lambda_{s}\frac{S_{M}}{Mf_{1}^2}\right)a^{\dag}, ~\quad
\end{eqnarray}
with
\begin{eqnarray*}
\nonumber
  C_{\lambda_{s}} &=&\cos(\lambda_{s}\tau),\qquad S_{\lambda_{s}}= \sin(\lambda_{s}\tau),\qquad f_{1}=\sqrt{2 (2a^{\dag}a+ 1)},
  \nonumber\\
  M&=&\sqrt{f_{1}^{2}+\lambda_{s}^{2}},~
C_{M}=\cos(M\tau),\qquad S_{M}= \sin(M\tau),\qquad \tau=\lambda t,
\nonumber\\
\lambda_{s}&=&\frac{\Omega_{s}}{\lambda},\quad  C_{M\lambda_s}=C_M-C_{\lambda_s}.
\end{eqnarray*}
In order to obtain the density operator of the atomic system $\rho_{aa}(t)$, one has to trace out the state of the field. The analytical solution is too complicated to be written through the text. However, as  a special case, the density operator of the atomic system in the excited state, namely we set  $\theta=0$, is given by,
\begin{eqnarray}\label{eq:a9}
  \rho_{aa}(t) &=&Tr_{f}\left[\rho(t)\right]=\left(
                                              \begin{array}{cccc}
                                                \rho_{1,1} & \rho_{1,2} & \rho_{1,3} & \rho_{1,4} \\
                                                \rho_{2,1} & \rho_{2,2} & \rho_{2,3} & \rho_{2,4} \\
                                                \rho_{3,1} & \rho_{3,2} & \rho_{3,3} & \rho_{3,4} \\
                                                \rho_{4,1} & \rho_{4,2} & \rho_{4,3} & \rho_{4,4} \\
                                              \end{array}
                                            \right),
\end{eqnarray}
where
\begin{eqnarray}\label{eq:a10}
  \rho_{1,1} &=&\sum_{n=0}^{+\infty}C_{n,n} \left(\left(C_{\lambda _s}+2 n \gamma _{n-1}\right){}^2+4 n^2 \alpha _{n-1}^2-4 n \alpha _{n-1} S_{\lambda _s}+S_{\lambda _s}^2\right),
\nonumber\\
\rho_{1,2} &=&\sum_{n=0}^{+\infty}\sqrt{n+1} \beta _n C_{n,n+1} \left(i C_{\lambda _s}-2 n \alpha _{n-1}+2 i n \gamma _{n-1}+S_{\lambda _s}\right),
\nonumber\\
\rho_{1,3} &=&
\sum_{n=0}^{+\infty}\sqrt{n+1} \beta _n C_{n,n+1} \left(i C_{\lambda _s}-2 n \alpha _{n-1}+2 i n \gamma _{n-1}+S_{\lambda _s}\right),
\nonumber\\
\rho_{1,4} &=&\sum_{n=0}^{+\infty}2 \sqrt{n+1} \sqrt{n+2} C_{n,n+2} \left(\gamma _{n+1}-i \alpha _{n+1}\right) \left(C_{\lambda _s}+2 n \left(\gamma _{n-1.}+i \alpha _{n-1.}\right)-i S_{\lambda _s}\right),
\nonumber\\
\rho_{2,2} &=&\sum_{n=0}^{+\infty}n \beta _{n-1.}^2 C_{n,n},
\quad
\rho_{2,3}=\sum_{n=0}^{+\infty}n \beta _{n-1}^2 C_{n,n},
\nonumber\\
\rho_{2,4} &=&
\sum_{n=0}^{+\infty}-2 n \sqrt{n+1} \beta _{n-1} C_{n,n+1} \left(\alpha _n+i \gamma _n\right),
\quad
\rho_{3,3} =\sum_{n=0}^{+\infty}n \beta _{n-1}^2 C_{n,n},
\nonumber\\
\rho_{3,4} &=&\sum_{n=0}^{+\infty}-2 n \sqrt{n+1} \beta _{n-1} C_{n,n+1} \left(\alpha _n+i \gamma _n\right),
\nonumber\\
\rho_{4,4} &=&\sum_{n=0}^{+\infty}
4 (n-1) n C_{n,n} \left(\alpha _{n-1}^2+\gamma _{n-1}^2\right),
\nonumber\\
\rho_{i,j}&=&\rho_{j,i}^{*},\qquad \qquad{i,j=1,2,3,4,i\neq j},
\end{eqnarray}
where
\begin{eqnarray*}
  \alpha_{n}=\left(\frac{S_{\lambda_{s}}}{f_{1}^2}-\lambda_{s}\frac{S_{M}}{Mf_{1}^2}\right),
  \quad
  \beta_{n}=\frac{S_{M}}{M},\quad
  \quad
  \gamma_{n}=\frac{C_{M\lambda_{s}}}{f_{1}^2}
  \quad
  \delta_{n}=\frac{C_{M\lambda_{s}}}{2}.
\end{eqnarray*}
Now, we have the details to discuss the entangling power of the AJCM model to generate entanglement between the atomic subsystems. Moreover its efficiency to perform quantum coding will be investigated.
\section{Entangling power}
This section aims to investigate the ability of the AJCM to generate quantum correlations between two atoms. The entangling power of the proposed model is assessed using the unitary operator \(\mathcal{U}_I(t)\), as described in \cite{RR18}.

\begin{equation}\label{eq:a11}
E_{\mathrm{p}}(U(t))=1-\frac{1}{d(d+1)}(\zeta_{1}+\zeta_{2}).
\end{equation}
where $d=4$, if the input state of the field is a coherent state $\rho^{in}=\sum_{\substack{m,n}}C_{\substack{m,n }}|m\rangle\langle n|$, $\zeta_{1}$ and $\zeta_{2}$ are given by

\begin{eqnarray*}
\zeta_{1}&=&\sum_{k_1, k_2=1}^{4}\left\{\left|tr\left(\sum_{i_1=1}^{4} \{U_{k_2 i_1}\sum_{\substack{m,n}}C_{\substack{m,n }}|m\rangle\langle n| U_{k_1 i_1}^{\dagger}\}\right)\right|^2\right\},
\\
\zeta_{2}&=&\sum_{k_1, k_2=1}^{4}\left\{\sum_{i_1, i_2=1}^{4}\left|tr\left(U_{k_2 i_2} \sum_{\substack{m,n}}C_{\substack{m,n }}|m\rangle\langle n| U_{k_1 i_1}^{\dagger}\right)\right|^2\right\}.
\end{eqnarray*}

\begin{figure}[h!]
 \begin{center}
\includegraphics[height=4.2cm,width=5.5cm]{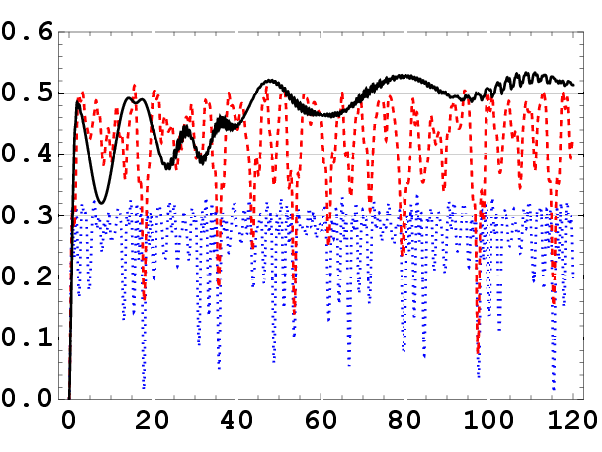}
 \put(-135,100){\small $(a)$}
 \put(-70,-5){$\tau$}
\put(-170,60){$E_{p}$}
\includegraphics[height=4.2cm,width=5.5cm]{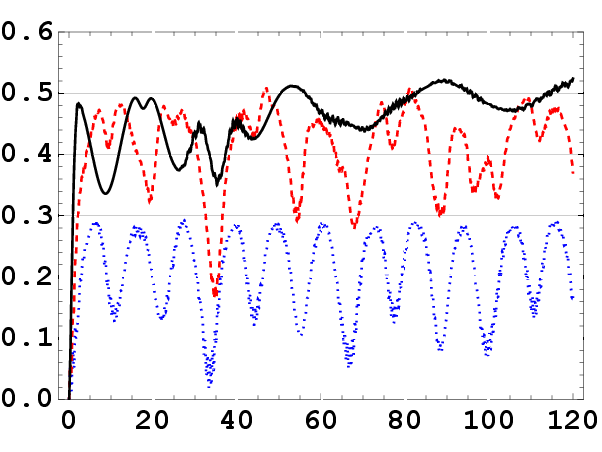}
 \put(-135,100){\small $(b)$}
  \put(-70,-5){$\tau$}
\includegraphics[height=4.2cm,width=5.5cm]{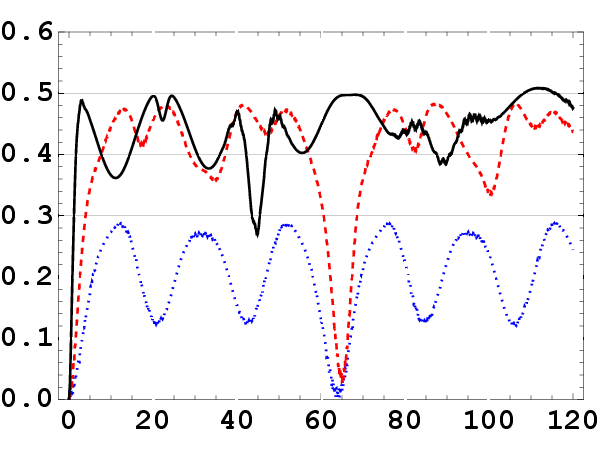}
 \put(-135,100){\small $(c)$}
  \put(-70,-5){$\tau$}
\caption{Evolution behaviour of the entangling power for two atoms when the field is initially in a coherent state, and for different mean photon number, $\bar{n}=0.1$ (dotted line), $\bar{n}=1$ (dashed line) and $\bar{n}=25$ (solid line)  and Ising parameter $\Omega_{s}$ where $\Omega_{s}=0$ (a), $\Omega_{s}=5\lambda$ (b), $\Omega_{s}=10\lambda$ (c)}
\label{F1}
\end{center}
\end{figure}
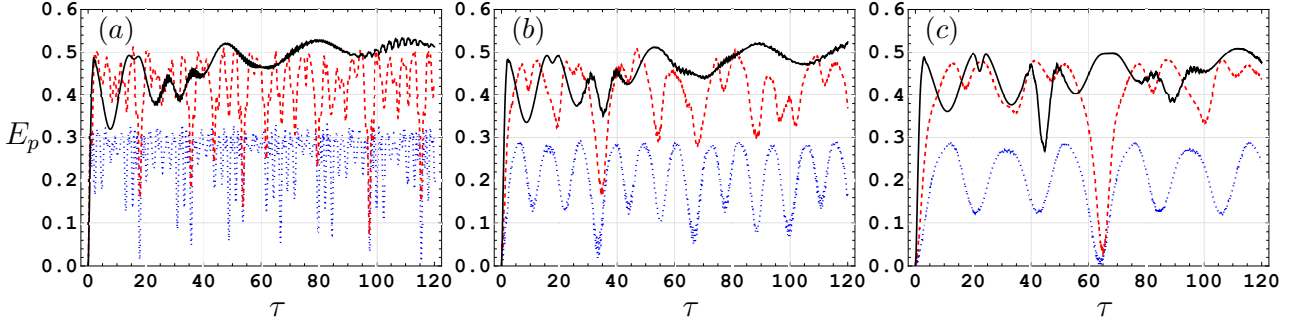

Fig.(\ref{F1}) illustrates the entangling power \(E_p\) and its dependence on interaction parameters. When the interaction is activated, the AJCM generates entanglement between the atomic subsystems, with \(E_p\) oscillating between upper and lower bounds influenced by the mean photon number \(\bar{n}\).
Fig. (\ref{F1}a) shows \(E_p\) without atom-atom interaction (\(\Omega_s = 0\)), where the mean photon number significantly affects the oscillations. At small \(\bar{n}\), \(E_p\) oscillates rapidly with upper bounds not exceeding 0.3. As \(\bar{n}\) increases, the upper bounds rise and oscillation amplitudes decrease, improving the lower bounds. For larger \(\bar{n}\), the upper bounds approach 0.5, and at \(\bar{n} = 25\), oscillations become fewer and smaller, indicating long-lived quantum correlations.
Fig. (\ref{F1}b) shows \(E_p\) with atom-atom interaction (\(\Omega_s = 5\lambda\)), where the behavior is similar to Fig. (\ref{F1}a) but with fewer, longer oscillations and more stable entanglement generation efficiency. The upper bounds are comparable to those in Fig. (\ref{F1}a).
Fig. (\ref{F1}c) shows \(E_p\) with increased interaction strength \(\Omega_s\), where oscillations have longer periodicity and smaller amplitudes, further stabilizing entanglement generation.
From Fig. (\ref{F1}), it can be concluded that increasing the mean photon number inside the cavity enhances the AJCM's ability to generate quantum correlations. The presence of atom-atom interaction reduces oscillation amplitudes, thereby increasing the lower bounds of AJCM efficiency. At higher interaction strengths, the AJCM's long-lived efficiency is predicted to improve.

\section{ Entanglement between atomic systems}
Now, it is important to quantify the amount of entanglement that may be generated between the two atoms. It is well known that the  concurrence  is a good measure of entanglement between two qubits system  \cite{RR34}. However for any two-qubit system defined by $\rho_{s}$, the mathematical form of the concurrence is defined as,
\begin{eqnarray}
  C &=& max\left(0,\sqrt{\lambda_{1}}-\sqrt{\lambda_{2}}-\sqrt{\lambda_{3}}-\sqrt{\lambda_{4}}\right),
\end{eqnarray}
where $\lambda_{1}\geq \lambda_{2}\geq\lambda_3\geq\lambda_{4}$ are the eigenvalues of the matrix $\varrho=\rho_{aa}(t)\left(\sigma_{y}\bigotimes\sigma_{y}\right)\rho_{aa}^{*}(t)\left(\sigma_{y}\bigotimes\sigma_{y}\right)$.
The range of the concurrence is from $0$ to $1$. For unentangled qubits
$C=0$, whereas $C=1$ for the maximally entangled qubits.

\begin{figure}[h!]
 \begin{center}
\includegraphics[height=4.2cm,width=5.5cm]{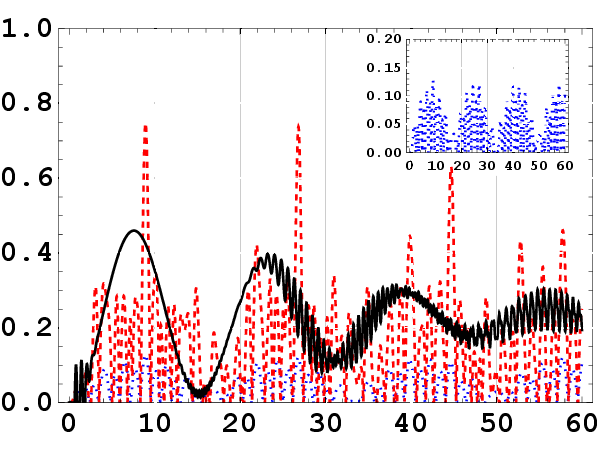}
 \put(-135,100){\small $(a)$}
 \put(-70,-5){$\tau$}
\put(-170,60){$C$}
\includegraphics[height=4.2cm,width=5.5cm]{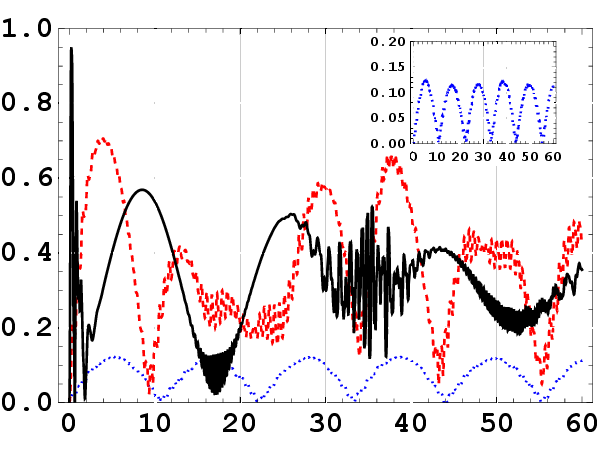}
 \put(-135,100){\small $(b)$}
  \put(-70,-5){$\tau$}
\includegraphics[height=4.2cm,width=5.5cm]{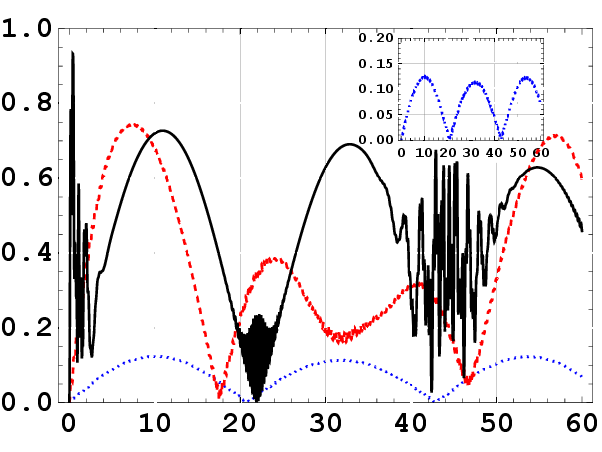}
 \put(-135,100){\small $(c)$}
  \put(-70,-5){$\tau$}
\caption{ The concurrence as a quantifier of entanglement, where we set $\theta=0$, $\Omega_s=0,5\lambda$, and $10\lambda$ for (a),(b) and (c) respectively. The dot dash and sold curves are evaluated at $\bar n=0.1,1$ and $25$ }
\label{FCC}
\end{center}
\end{figure}
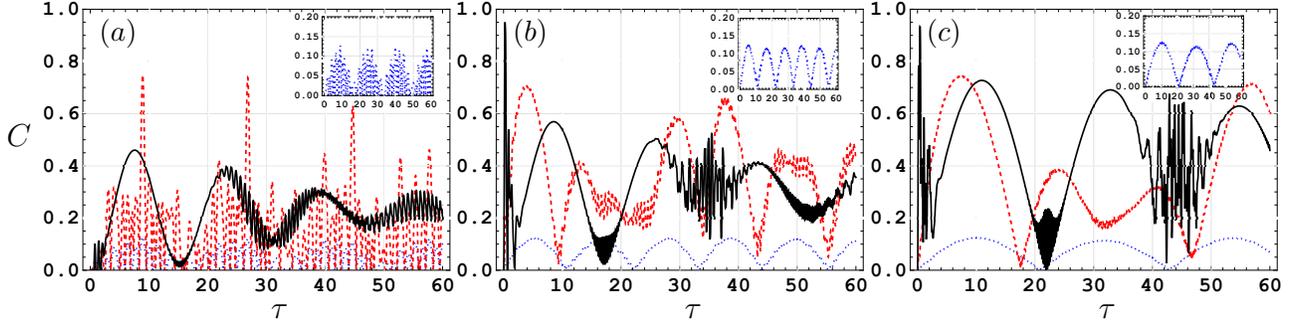

In Fig.(\ref{FCC}), we investigate the behavior of  the concurrence  as a quantifier of the amount of entanglement that contained in the state $\rho_{aa}(t)$ in the presence and absence of the atom-atom interaction's strength. It is observed that, at $\Omega_s=0$, an entanglement is generated between the two atoms, as soon as the interaction of the atomic system with the cavity mode is switched on. Moreover, the fast oscillations behavior, as well as the collapse phenomena  are predicted at small values of the mean photon number. However, the phenomenon of the sudden death/birth phenomenon of entanglement is observed periodically as it is shown in Fig.(\ref{FCC}a), the shortest periodic time is observed as one decreases the mean photon number inside the cavity.
For the non-zero value of the atomic interaction strength, $\Omega_s$, the concurrence's behavior is displayed in Figs.(\ref{FCC}b) and (\ref{FCC}c), where we set $\Omega_s=5\lambda, 10\lambda$, respectively. It is clear that, the concurrence vanishes periodical at small values of the mean photon number.  As one increase $\bar n$, the long-lived entanglement is observed, where the periodically vanishing time increases.  Moreover, the largest amount of entanglement is observed as one increase  $\bar n$.

From Fig.(\ref{FCC}), one can say that it is possible to increase the amount of quantum correlation between the two atoms by increasing the mean photon number  inside the cavity and the interaction strength between the two atoms. Moreover,  the interaction parameters could be used as controllers  to obtain a long-lived entanglement.
\begin{figure}[h!]
 \begin{center}
\includegraphics[height=4.2cm,width=5.5cm]{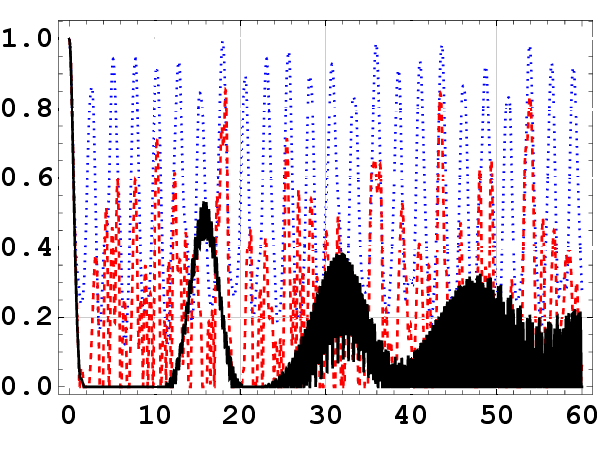}
 \put(-135,100){\small $(a)$}
 \put(-70,-5){$\tau$}
\put(-170,60){$C$}
\includegraphics[height=4.2cm,width=5.5cm]{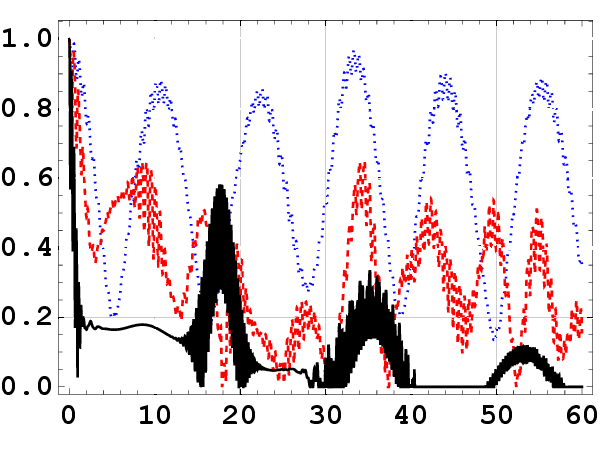}
 \put(-135,100){\small $(b)$}
  \put(-70,-5){$\tau$}
\includegraphics[height=4.2cm,width=5.5cm]{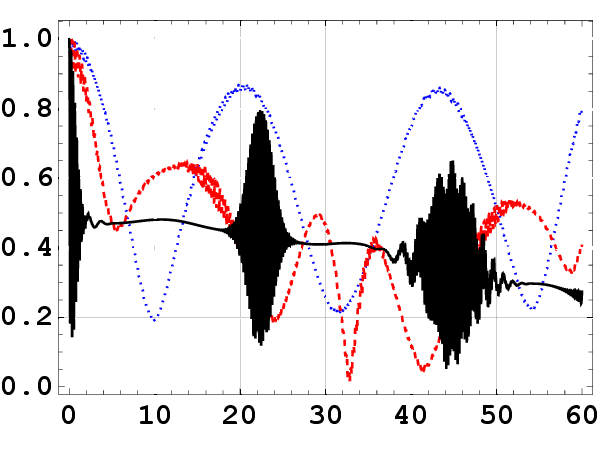}
 \put(-135,100){\small $(c)$}
  \put(-70,-5){$\tau$}
\caption{ The concurrence as a quantifier of entanglement, where we set $\theta=\pi/4$, $\Omega_s=0,5\lambda$, and $10\lambda$ for (a),(b) and (c) respectively. The dot dash and sold curves are evaluated at $\bar n=0.1,1$ and $25$ }
\label{FCC20}
\end{center}
\end{figure}
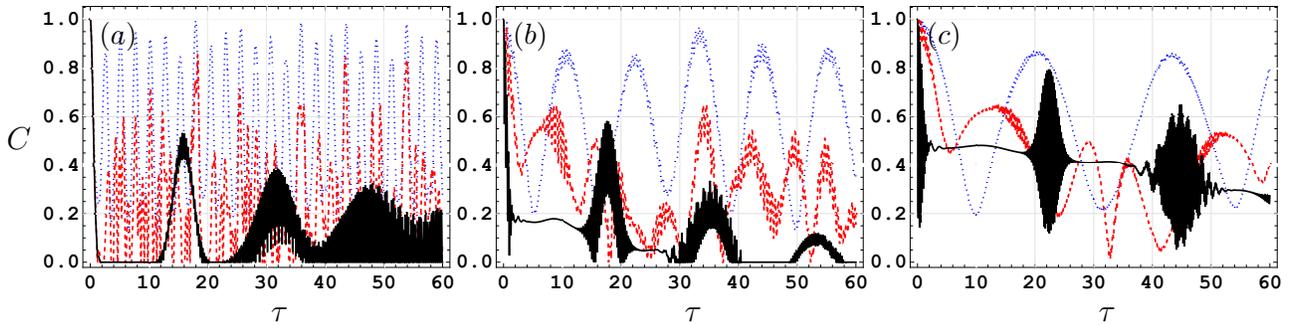

In Fig.(\ref{FCC20}), we investigate the effect of the AJCM on the initial amount of entanglement in the presence/absence of the atomic interaction'strength.  It is clear that, at small values of $\bar n$, the concurrence oscillates between its  maximum values and non-zero bounds.  However, as one increase the mean photon number inside the cavity, the possibility that the field interacts with each atom increases and consequently the initial amount of entanglement of the atomic system decreases. As it is displayed from Fig.(\ref{FCC20}a), where $\Omega_s=0$, the phenomenon of the sudden changes of the concurrence increases  at large values of $\bar n$. Moreover, concurrence vanishes periodically, where these periods increase one increases the mean photon number.  In Figs.(\ref{FCC20}b) and (\ref{FCC20}c), we consider non-zero values of $\Omega_s$. The observed behavior of the concurrence shows  a long  periodical time, the lower bounds of entanglement are improved, long-lived entanglement, and collapses/revivals behaviors of entanglement.
\section{ Atomic-channel capacity}
As an important application  of the generated channel between the two atoms, its ability to code information. Therefore investigating the impact of the interaction parameters on  the capacity behavior of the atomic system state, is one aim of this study. It is well known that the channel capacity quantifies the amount of information that may be coded on a quantum state. For any two qubit state $\rho$, its channel capacity is defined as \cite{RR35},
\begin{eqnarray}\label{eq:a12}
  \chi(t) &=&S(\overline{\rho^{*}})-S(\rho_{aa}),
\end{eqnarray}
where $S(\rho)=-tr\left(\rho\log_{2}(\rho)\right)$ the von Neumann entropy of density matrix and  $\overline{\rho^{*}}$ represents  the average state in which the information has been encoded. By using  the set of mutually orthogonal unitary transformations, $U_{i}$, one can write $\overline{\rho^{*}}$ as\cite{RR36},
\begin{eqnarray}\label{eq:a13}
  \overline{\rho^{*}} &=&\frac{1}{4}\sum_{i=0}^{3}(U_{i}\otimes I)\rho_{aa}(U_{i}\otimes I),
\end{eqnarray}
where,
\begin{eqnarray*}\label{eq:a14}
  U_{0}&=&\left(
            \begin{array}{cc}
              1 & 0 \\
              0 & 1 \\
            \end{array}
          \right),\quad
U_{1}=\left(
            \begin{array}{cc}
              0 & 1 \\
              1 & 0 \\
            \end{array}
          \right),
\quad
U_{2}=\left(
            \begin{array}{cc}
              -1 & 0 \\
              0 & 1 \\
            \end{array}
          \right),
\quad
U_{3}=\left(
            \begin{array}{cc}
              0 & 1 \\
              -1 & 0 \\
            \end{array}
          \right).
\end{eqnarray*}

\begin{figure}[h!]
 \begin{center}
\includegraphics[height=4.2cm,width=5.5cm]{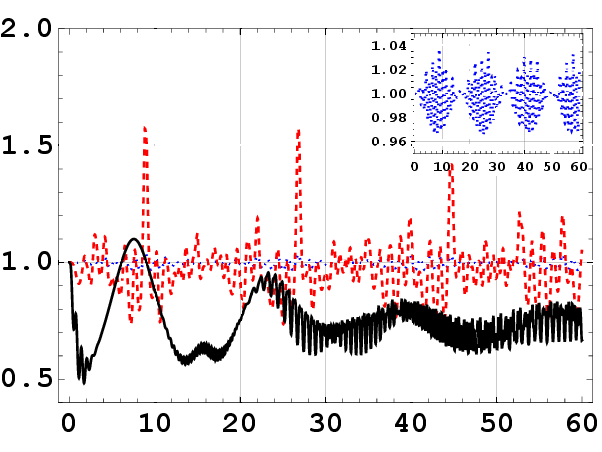}
 \put(-135,100){\small $(a)$}
 \put(-70,-5){$\tau$}
\put(-180,60){$\chi(t)$}
\includegraphics[height=4.2cm,width=5.5cm]{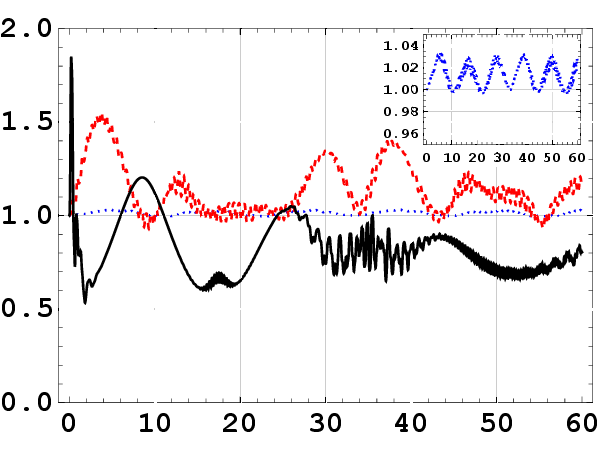}
 \put(-135,100){\small $(b)$}
  \put(-70,-5){$\tau$}
\includegraphics[height=4.2cm,width=5.5cm]{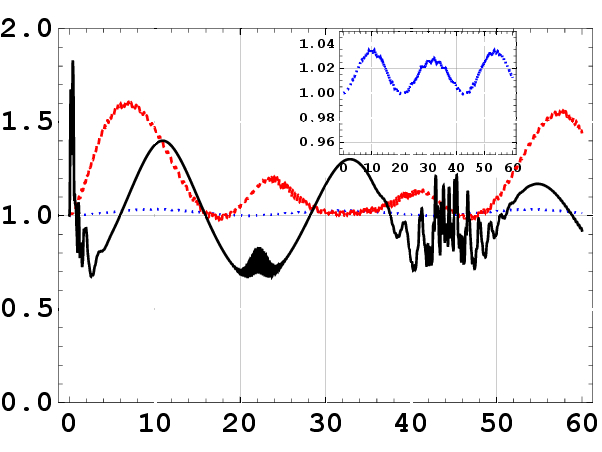}
 \put(-135,100){\small $(c)$}
  \put(-70,-5){$\tau$}
\caption{The channel  capacity behavior of the atomic system, which is initially prepared in a separable state, $\rho_{aa}=\ket{ee}$. The  dot,dash and solid curves are evaluated at $\bar n=0.1,1,25$, respectively, where we set  $\Omega_{s}=0,5\lambda,10\lambda$ in (a),(b) and (c), respectively.}
\label{F2D}
\end{center}
\end{figure}

In Fig.(\ref{F2D}), we examine the ability  of the atomic channel to  encode information by investigating its capacity, where it is assumed that the atoms are initially prepared in a separable state .  It is clear that, the behavior of the capacity function $\chi(t)$ is similar to that shown for the concurrence. However, in the absence of the  atom-atom interaction and small values of the mean photon number inside the cavity, the oscillations and collapses behavior of the $\chi(t)$ function are observed. As one increase the mean photon number, the  stable behavior of the channel capacity is predicted, where  the periodical time of the oscillation increases and its amplitudes decrease. Therefore, the upper bounds of the channel capacity are improved.

\begin{figure}[h!]
 \begin{center}
\includegraphics[height=4.2cm,width=5.5cm]{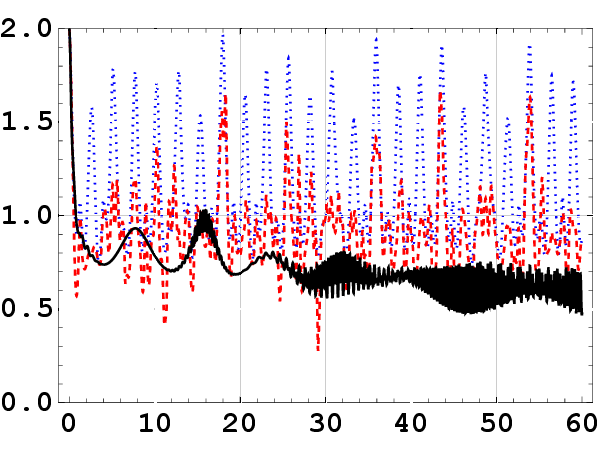}
 \put(-135,100){\small $(a)$}
 \put(-70,-5){$\tau$}
\put(-180,60){$\chi(t)$}
\includegraphics[height=4.2cm,width=5.5cm]{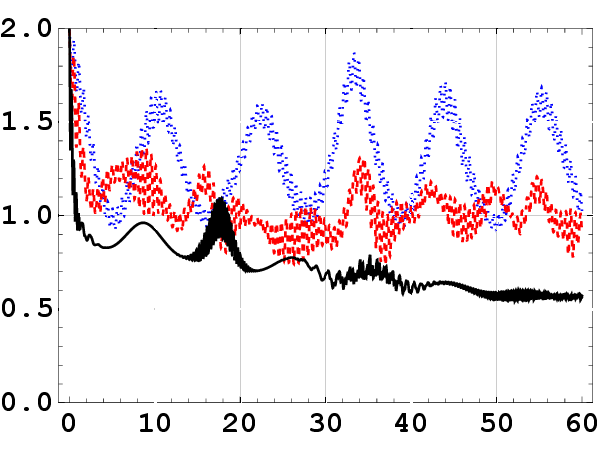}
 \put(-135,100){\small $(b)$}
  \put(-70,-5){$\tau$}
\includegraphics[height=4.2cm,width=5.5cm]{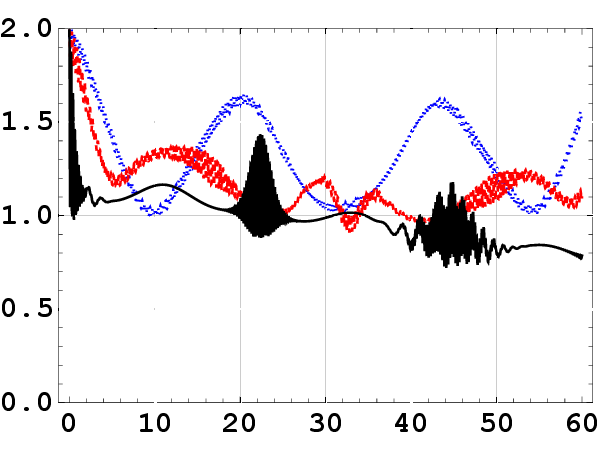}
 \put(-135,100){\small $(c)$}
  \put(-70,-5){$\tau$}
\caption{ The same as Fig.(\ref{F2D}), but the atomic system is initially prepared in a maximum entangled state.}
\label{F2D00}
\end{center}
\end{figure}
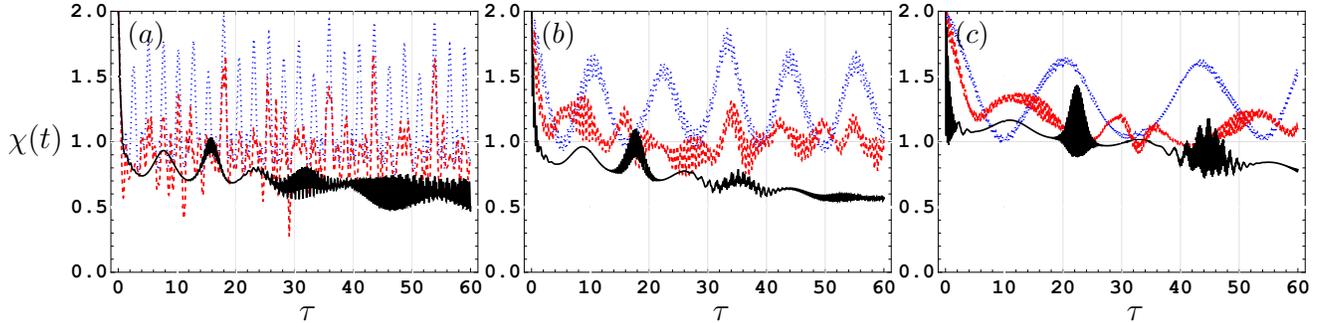
The impact  of the AJCM on the channel capacity of the atomic state that has initially prepared in a maximum entangled state is described in Fig.(\ref{F2D00}).  In general, the channel capacity $\chi(t)$ increases as one decreases the mean photon number, where the atomic state has a strong coherence. However, Fig.(\ref{F2D00}a) shows fast oscillations, where the amplitudes of these oscillations decrease as one increase the mean photon number, and consequently its lower bounds exceed the classical ones. For non-zero strength of the atomic interaction, the large periodical behavior is observed, and the lower bounds of $\chi(t)$ are improved.

From Figs.(\ref{F2D}) and (\ref{F2D00}) it is observed that, the channel capacity depends on the initial state settings of the atomic system, where by  starting from maximum entangled state of the atomic system, the lower bounds of the channel capacity and its periodical time increase.

\section{ Conclusions}
This study investigates the capability of the Anti-Jaynes-Cummings Model (AJCM) to generate entanglement between two atoms, initially prepared in either a separable state or a maximally entangled state. An additional Ising-type interaction between the atoms is also considered. The degree of entanglement between the atoms is quantified using concurrence, and the potential of the atomic state as a quantum channel for encoding information is analyzed.

The results reveal that the AJCM can generate entanglement between the atomic subsystems immediately after the interaction is initiated. The entangling power function oscillates between lower and upper bounds, which are influenced by the interaction parameters, with the atom-atom interaction strength playing a critical role in enhancing these bounds. At low mean photon numbers within the cavity, the system exhibits oscillations along with collapse and revival phenomena. However, as the mean photon number increases, both the frequency and amplitude of oscillations decrease, leading to improved lower and upper bounds of the entangling power. Additionally, as the oscillation period lengthens over extended interaction times, the AJCM demonstrates an increased long-term entangling capability.

The amount of entanglement is quantified using concurrence in both the presence and absence of atom-atom interaction. The predicted behavior of concurrence indicates that entanglement between the two atoms depends on the initial state of the atomic system and the interaction parameters. When starting from a separable atomic system, entanglement periodically vanishes at low mean photon numbers, with the period of this phenomenon increasing as the atom-atom interaction strength grows.
In the absence of atom-atom interaction, the entanglement decreases as the mean photon number inside the cavity increases. However, for non-zero interaction strength between the atoms, the lower bounds of entanglement improve. When the atomic system is initially prepared in a maximally entangled state (MES), both the periodicity and lower bounds of entanglement improve with higher mean photon numbers and increased atom-atom interaction strength.

The potential use of the final atomic state for encoding information is a key application in quantum information processing. To explore this, we analyzed the behavior of the quantum capacity of the entangled state generated between the two atoms. Our findings show that interaction parameters can serve as control mechanisms to enhance the atomic state's capacity, thereby improving its information encoding capabilities. In the absence of atomic interaction strength, higher capacity is observed at low mean photon numbers. However, when the atomic interaction strength is non-zero, the channel capacity is further enhanced.

In summary, the AJCM efficiently generates entanglement between two atoms initially in separable states, enabling information encoding. Interaction parameters act as controls to enhance the state's efficiency. Starting from a maximally entangled state (MES), the lower bounds and periodicity of entanglement improve with non-zero atomic interaction strength.

\section*{Acknowledgements} We sincerely thank the referees for their valuable remarks, which have helped us enhance the revised version of our work.

\vspace{0.5cm}
{\bf Declaration of Interest:} The authors declare that they have no conflict of interest.\\
{\bf Data availability statement:} No data was used for the research described in the article.\\
{\bf Author contributions:} The authors confirm that all of  them have  contributed
equally in preparing this manuscript.

\end{document}